\documentclass[pra,superscriptaddress,showkeys,amsmath,amssymb,twocolumn]{revtex4-2}
\usepackage[utf8]{inputenc}
\usepackage[T1]{fontenc}
\usepackage{lmodern}
\usepackage{graphicx}
\usepackage{color}
\usepackage{dcolumn}
\usepackage{bm}

\usepackage[colorlinks,bookmarks=true,citecolor=blue,linkcolor=blue,urlcolor=blue]{hyperref}

\RequirePackage[normalem]{ulem}
\RequirePackage{color}

\newcommand{\mathsout}[1]{%
  \mathchoice
    {\text{\sout{$\displaystyle #1$}}}
    {\text{\sout{$\textstyle #1$}}}
    {\text{\sout{$\scriptstyle #1$}}}
    {\text{\sout{$\scriptscriptstyle #1$}}}%
}
\DeclareRobustCommand{\old}[1]{%
  \ifmmode
    \textcolor{red}{\mathsout{#1}}%
  \else
    \textcolor{red}{\sout{#1}}%
  \fi
}

\begin{document}
\preprint{AIP/123-QED}

\title{Information Entropy as a Sensitive Diagnostic of Relaxation Dynamics in a Quasiperiodic Mott Insulator}

\author{N.~D.~Chavda}
\affiliation{Department of Applied Physics, Faculty of Technology and Engineering,The Maharaja Sayajirao University of Baroda, Vadodara-390001, India.}
\author{Antônio~M.~S.~Macêdo}
\affiliation{ Laboratório de Física Teórica e Computacional, Departamento de Física,
Universidade Federal de Pernambuco, 50670-901 Recife, Pernambuco, Brazil}
\author{Robin~P~Sagar}
\affiliation{Departamento de Química, Universidad Autónoma Metropolitana -
Iztapalapa, Rafael Atlixco 186, CDMX, 09340, Ciudad de México,
México.}
\author{Barnali~Chakrabarti}
\email{barnali.chakrabarti@ufpe.br}
\affiliation{ Laboratório de Física Teórica e Computacional, Departamento de Física,
Universidade Federal de Pernambuco, 50670-901 Recife, Pernambuco, Brazil}

\date{\today} 

\begin{abstract}
We study the nonequilibrium dynamics of a bosonic Mott insulator following a sudden quench from a periodic to a quasiperiodic lattice. Using ab initio multiconfigurational (MCTDHB) continuum many-body dynamics, we characterize the evolution through momentum distributions, fragmentation, correlation functions, and information entropy. We find that melting of Mott correlations does not necessarily imply relaxation of the many-body state. In the weakly interacting regime, the initial Mott-like correlation undergo recurrent melting and partial recovery, while the information entropy displays large, intermittent fluctuations within the simulated time window. In contrast, a strongly interacting and fully fragmented Mott insulator retains its characteristic Mott correlations, up to partial distortions, while the many-body information entropy approaches a well-defined stationary regime plateau with comparatively small fluctuations. These results indicate that Mott-correlation dynamics and many-body relaxation can be dissociated in this finite system and support the use of information entropy, evaluated in the multiconfigurational representation, as a sensitive complementary diagnostic of nonequilibrium relaxation.

\end{abstract}


\maketitle

\section{Introduction}
\label{sec:intro}
The interplay between disorder and interactions in quantum many-body systems gives rise to a rich variety of insulating and localized phases and remains a central problem in condensed-matter and quantum-gas physics~\cite{Giamarchi:1988,Fisher:1989}. A paradigmatic example is the Mott insulator (MI), which emerges at commensurate filling when interactions suppress density fluctuations and establish strong local correlations. In the presence of disorder, these interaction-induced correlations compete with localization, leading to distinct insulating regimes such as the Bose-glass phase~\cite{Giamarchi:1988,Fisher:1989,Fallani:2005,Palencia:2019,Pasienski:2010,Yu:2012,Ronen:2012}. The interplay between disorder and interactions has been extensively explored in ultracold gases, including disorder-induced superfluid--Bose-glass transitions~\cite{Marino_2026,Tanzi_2013,Chiara1,Meldgin:2016,Palencia:2010} and unconventional modifications of Mott phases~\cite{Palencia:2019,Palencia:2020,BarnaliPRB}.

Quasiperiodic optical lattices provide a particularly controlled realization of correlated disorder, combining long-range spatial structure with broken translational
symmetry~\cite{Aubry:1980,Roati:2008}. High-precision optical-lattice experiments have enabled systematic studies of localization and interaction effects in one-dimensional bichromatic lattices~\cite{Roati:2008,Bordia:2017,Lueschen:2017,Nakajima:2021,Gautier:2021,yao2024}. Such systems provide a versatile setting for studying the competition between localization induced by quasiperiodicity and many-body correlations generated by interactions. Beyond localization, quasiperiodic systems have revealed topological structures~\cite{Lang:2012,Dareau:2017,Baboux:2017}, fractal properties~\cite{Yuan:2000,Tanese:2014}, and unusual localization behavior~\cite{Roati:2008,Molignini_2025,Molignini:2025}. The equilibrium properties of interacting bosons in quasiperiodic potentials have also been investigated in one and higher dimensions~\cite{Zhu,Roux,Gautier:2021,Dalfovo:2009,Dalfovo:2011,Dalfovo:2019}.

While equilibrium properties of interacting bosons in disordered and quasiperiodic potentials have been extensively studied, their nonequilibrium response following a quantum quench can involve several coupled forms of dynamics~\cite{Berges_2004,Langen_2016,Rigol,Langen_2015,Pr_fer_2018,Erne_2018,Garc_a_Orozco_2022}. A quench modifies the spatial density, coherence, and interparticle correlations, but the evolution of these observables does not necessarily provide a direct measure of how the information content of the many-body state is redistributed. Information-theoretic
measures have therefore been widely used to characterize the
information content, correlations, and organization of quantum
many-body states~\cite{PhysRevA.109.063308,Paolo,rhombik_quantumreports,rhombik_jpb,rhombik_pre}. An important question is therefore how correlation dynamics and information dynamics are related following a nonequilibrium quench.

To address this question, we consider a bosonic Mott insulator subjected to a sudden quench from a periodic to a quasiperiodic lattice. Previous studies have used many-body entropies, fidelity, and correlation functions to investigate the nonequilibrium dynamics of interacting bosons in optical lattices~\cite{barnali_axel,rhombik_pra,rhombik_quantumreports,rhombik_pre,Marino_2026,Marino2026preprint}. Building on these investigations, we compare weakly and strongly interacting Mott states under the same quench protocol, examining how the interaction strength modifies the natural-occupation spectrum of the initial state, the relevant energy scales, and the energy injected by the quench. At weak interaction, the initial state is partially fragmented, with nearly but not exactly uniform natural occupations, and the quench induces pronounced dynamics characterized by recurrent melting and partial recovery of the Mott-like correlation structure. At strong interaction, the initial state is fully fragmented, while its characteristic Mott correlations remain largely preserved despite partial distortions during the evolution. This comparison reveals that the evolution of Mott correlations and the redistribution of many-body information can follow qualitatively different dynamical pathways.

We investigate whether the contrasting correlation dynamics are accompanied by corresponding changes in the information content of the evolving many-body state. To this end, we combine conventional observables and one- and two-body correlation functions with information-entropy measures constructed from the many-body wave function and the evolving spatial density. This combined analysis tracks the evolution of Mott correlations alongside the redistribution of information in the many-body state. The dynamics are obtained using the multiconfigurational time-dependent Hartree method for bosons (MCTDHB)~\cite{Streltsov:2006,Streltsov:2007,Alon:2007,Alon:2008,Lode:2016,Lode:2020}, as implemented in the MCTDH-X package~\cite{MCTDHX}, for $N=9$ bosons in $S=9$ lattice sites with hard-wall boundaries.

We find that correlation dynamics and information relaxation follow distinct pathways. In the weakly interacting regime, the characteristic Mott correlations undergo recurrent melting and partial recovery after the quench, accompanied by pronounced oscillations in conventional observables, while the information-entropy measures exhibit large, intermittent fluctuations throughout the simulated time window. In contrast, strongly interacting, fully fragmented Mott states retain their characteristic correlations at the sampled times, despite partial distortions, while the information-entropy measures reveal substantial redistribution and, for sufficiently strong quasiperiodic quenches, a three-stage relaxation toward a regime of comparatively small fluctuations around well-defined mean values. These results establish that Mott-correlation melting and many-body information relaxation are distinct dynamical processes: \emph{correlation melting can occur without clear information relaxation, whereas comparatively relaxed information dynamics, within the employed representation, can emerge without the destruction of Mott order.}

We investigate whether these differences in correlation dynamics are accompanied by corresponding changes in the information content of the evolving many-body state. To this end, we combine conventional observables and one- and two-body correlation functions with information-entropy measures constructed from the many-body wave function and the evolving spatial density. This combination allows us to examine the evolution of conventional Mott correlations together with the redistribution and statistical relaxation of information in the evolving many-body state. The dynamics are obtained using the multiconfigurational time-dependent Hartree method for bosons (MCTDHB)~\cite{Streltsov:2006,Streltsov:2007,Alon:2007,Alon:2008,Lode:2016,Lode:2020}, as implemented in the MCTDH-X package~\cite{MCTDHX}, for $N=9$ bosons in $S=9$ lattice sites with hard-wall boundaries.

The paper is organized as follows. Section~II introduces the Hamiltonian and quench protocol, while Sec.~III describes the MCTDHB approach and numerical implementation. Section~IV defines the dynamical and information-entropy observables used in the analysis. Section~V characterizes the initial Mott states and their one- and two-body correlation properties. Section~VI presents the nonequilibrium dynamics of the weakly interacting Mott state, including the evolution of conventional observables, Mott correlations, and information-entropy measures. Section~VII examines the corresponding dynamics in the strongly interacting regime and analyzes the distinction between the persistence of Mott correlations and many-body information relaxation. Section~VIII summarizes our main conclusions.

\section{System and Protocol}

The dynamics of $N$ interacting bosons of mass $m$ confined in a one-dimensional optical lattice are governed by the time-dependent Schrödinger equation,

\begin{equation}
    \hat{H} \Psi = i\hbar \frac{\partial \Psi}{\partial t}.
\end{equation}
The total $N$-body Hamiltonian reads

\begin{equation}
\hat{H}(x_1, x_2, \dots, x_N) = \sum_{i=1}^{N} \hat{h}(x_i) + \sum_{i<j=1}^{N} \hat{W}(x_i - x_j),
\label{propagation_eq}
\end{equation}
where the one-body part $\hat{h}(x) = \hat{T}(x) + \hat{V}(x)$ consists of the kinetic energy operator $\hat{T}(x) = -\frac{\hbar^2}{2m}\frac{\partial^2}{\partial x^2}$ and the external potential $\hat{V}(x)$.

We express the Hamiltonian in dimensionless units by scaling lengths
with $\bar L=1/k_p$ and energies with
$\bar E=\hbar^2/(m\bar L^2)=2E_r$, where
$E_r=\hbar^2k_p^2/(2m)$ is the recoil energy of the primary optical
lattice. The corresponding time unit is $\hbar/\bar E$ (see
Appendix~A). In these units, the bichromatic optical lattice potential
is
\begin{equation}
    V(x)=V_p\sin^2(x)+V_d\sin^2(rx),
\end{equation}
where $V_p$ and $V_d$ are the dimensionless depths of the primary and
secondary lattices, respectively, and $r=k_d/k_p$ is the ratio of
their wave vectors. The experimental wavelengths
$\lambda_p=532.2\,\mathrm{nm}$ and
$\lambda_d\simeq444.5\,\mathrm{nm}$ give
$r\simeq1.197215$. This ratio is a finite-precision approximation to
an irrational number, as appropriate for a quasiperiodic lattice
\cite{Roati:2008}. The primary-lattice spacing is $\pi$ in the
dimensionless coordinate $x$.

The bosons interact through a short-range two-body contact potential,
\begin{equation}
    W(x_i-x_j)=g_0\delta(x_i-x_j),
\end{equation}
where $g_0=g_{1D}/(\bar E\bar L)>0$ is the dimensionless interaction
strength and $g_{1D}$ is the one-dimensional contact coupling, with
dimensions of energy times length. The parameter $g_0$ should not be
identified with the on-site interaction energy $U$ of the
Bose--Hubbard model. All quantities below are expressed in the
dimensionless units defined above.

We consider $N=9$ bosons in a finite primary lattice comprising $S=9$
sites, with the lattice depth fixed at $V_p=10$ and hard-wall boundaries
at $x=\pm9\pi/2$. The site spacing is defined by the distance between
adjacent maxima of the primary sinusoidal potential. At this
commensurate filling, increasing $g_0$ drives the system from a
delocalized, superfluid-like state through a weakly interacting
Mott-like regime to a strongly interacting, fully fragmented
Mott-like state. Because the system is finite, these changes represent
a crossover rather than a sharp phase transition, allowing us to
examine how increasing interactions modify localization, fragmentation,
and Mott correlations.

The relative phase between the two lattices is fixed at zero, as
specified by the potential above; no averaging over this phase is
performed. With the symmetric hard-wall boundaries, the potential is
parity symmetric, and the dynamics preserves the parity sector of the
initial state. Our results therefore correspond to this specific phase
and boundary realization. The beat length of the bichromatic potential,
$\pi/(r-1)$, is approximately $5.1$ primary-lattice sites, so the
system spans fewer than two beat lengths.

The quench protocol is the same for all cases. We first prepare the ground state in the primary lattice. At $t>0$, the system is evolved under the Hamiltonian with a secondary lattice of depth $V_d$ suddenly switched on, driving the system far from equilibrium.

\section{Methods}
\label{app:MCTDHX}

To simulate the full many-body dynamics, we employ the multi configurational time-dependent Hartree method for bosons (MCTDHB)~\cite{Streltsov:2006,Streltsov:2007,Alon:2007,Alon:2008,Lode:2016,Fasshauer:2016,Lode:2020} via the MCTDH-X package~\cite{Lin:2020,MCTDHX}. MCTDH-X solves the time-dependent Schrödinger equation using a variationally optimized, time-dependent basis. The many-body wave function is expanded in $M$
single-particle functions (orbitals). Both the coefficients and the orbitals evolve self-consistently with time. Propagation in imaginary time results to the ground state and real time propagation provides non equilibrium dynamics.

The wave function of $N$ interacting bosons is expressed as a linear combination of time-dependent permanents,
\begin{equation}
    |\Psi(t)\rangle = \sum_{\mathbf{n}} C_{\mathbf{n}}(t) |\mathbf{n}; t\rangle,
    \label{many_body_wf}
\end{equation}
where $\mathbf{n} = (n_1, n_2, \dots, n_M)$ specifies the occupation numbers of the $M$ orbitals, constrained by $\sum_{k=1}^{M} n_k = N$.
Each permanent is constructed from the time-dependent creation operators $\hat{b}_k^\dagger(t)$ acting on the vacuum $|0\rangle$,
\begin{equation}
    |\mathbf{n}; t\rangle = \prod_{k=1}^{M} \frac{[\hat{b}_k^\dagger(t)]^{n_k}}{\sqrt{n_k!}} |0\rangle,
    \label{many_body_wf_2}
\end{equation}
where $\hat{b}_k^\dagger(t)$ creates a boson in the $k$th time-dependent orbital $\psi_k(x;t)$.
The total number of many-body configurations included in the expansion is
\begin{equation}
    N_{\mathrm{conf}} = \binom{N + M - 1}{N}.
\end{equation}

Applying the variational principle to the ansatz in Eq.~(\ref{many_body_wf}) yields coupled equations of motion for the coefficients  $C_{\mathbf{n}}(t)$ and the orbitals. Stationarity of the action with respect to independent variations produces two self-consistent equations, solved simultaneously at each time step, ensuring that both orbitals and coefficients remain variationally optimal for the instantaneous many-body state~\cite{TDVM81,variational1,variational3,variational4}.

For $M=1$, MCTDH-X reduces to the mean-field Gross–Pitaevskii description, while in the limit $M \rightarrow \infty$ it becomes exact, spanning the full
$N$-particle Hilbert space. In practice, $M$ is truncated for computational feasibility and convergence requires that (i) the occupation of the highest orbital is negligible and (ii) all relevant observables—density, correlations, and dynamical measures—are insensitive to further increases in $M$.

We solve the many-body dynamics using MCTDHB with $M=10$ time-dependent orbitals. This choice allows us to describe the correlated dynamics beyond a fixed-orbital or mean-field approximation. The spatial grid and other numerical parameters are specified in Appendix~A.

\section{Quantities of Interest}
\label{sec:quantity}

In this section, we define the physical quantities used to characterize the system dynamics, including the one-body density, the reduced one-body density matrix (RDM) and its eigenvalues, the Glauber correlation functions, and the one-, two-, and many-body information entropies.

\vspace{0.3cm}
\noindent
\emph{One-body density, reduced density matrix, natural occupations, and fragmentation.}\\[0.1cm]

To probe the spatial distribution of the bosons, we compute the one-body density,
\begin{equation}
\rho(x,t) = \langle \Psi(t) | \hat{\Psi}^{\dagger}(x) \hat{\Psi}(x) | \Psi(t) \rangle.
\end{equation}
The reduced one-body density matrix is defined as
\begin{equation}
\rho^{(1)}(x,x';t) = \langle \Psi(t) | \hat{\Psi}^{\dagger}(x) \hat{\Psi}(x') | \Psi(t) \rangle.
\end{equation}
From $\rho^{(1)}(x,x';t)$ one obtains the natural orbitals $\phi_j(x;t)$ and their respective occupations $\rho_j(t)$ through the eigenvalue equation
\begin{equation}
\rho^{(1)}(x,x';t) = \sum_{j=1}^{M} \rho_j(t) \, \phi_j^*(x;t)\,\phi_j(x';t),
\end{equation}
where the eigenvalues satisfy $\sum_{j=1}^{M} \rho_j(t) = N$; we denote by $n_j(t)=\rho_j(t)/N$ the normalized occupations, with $\sum_j n_j(t)=1$.
The natural orbitals are ordered in decreasing occupation, $n_1 \ge n_2 \ge \dots \ge n_M$.

Fragmentation, which MCTDHB captures by construction, occurs when more than one natural orbital is macroscopically occupied; in a finite system, ``macroscopic'' is used here in the operational sense of an appreciable fraction of the particles. The occupations $\{n_j(t)\}$ quantify dynamical fragmentation during non equilibrium evolution. In a fully condensed state, nearly all particles occupy a single orbital ($n_1 \approx 1$), while significant occupation of $k$ orbitals indicates $k$-fold fragmentation.

The degree of fragmentation is quantified by
\begin{equation}
\mathcal{F}(t)=1-n_1(t),
\end{equation}
where \(n_1(t)\) is the largest normalized natural occupation. Thus, \(\mathcal{F}=0\) corresponds to a fully condensed state, while larger values indicate stronger fragmentation.
Fragmentation thus measures the deviation of the many-body state from a mean-field single-orbital description.

\vspace{0.3cm}
\noindent
\emph{One-body and two-body correlation functions.} \\[0.1cm]

The coherence properties and many-body correlations are quantified using the normalized first- and second-order Glauber correlation functions,
\begin{align}
    g^{(1)}(x,x';t) &= \frac{\rho^{(1)}(x,x';t)}{\sqrt{\rho(x,t)\rho(x',t)}}, \\
    g^{(2)}(x,x';t) &= \frac{\rho^{(2)}(x,x';t)}{\rho(x,t)\rho(x',t)},
\end{align}
where the diagonal two-body density  is defined as
\begin{equation}
\rho^{(2)}(x,x';t) = \langle \Psi(t) |
\hat{\Psi}^{\dagger}(x)\hat{\Psi}^{\dagger}(x')\hat{\Psi}(x')\hat{\Psi}(x)
| \Psi(t) \rangle.
\end{equation}
Here, \(x\) and \(x'\) denote the positions of two particles, rather than off-diagonal coordinates.
Note that $g^{(1)}(x,x;t)=1$ identically wherever $\rho(x,t)>0$. 
Thus, the information contained in $|g^{(1)}|^2$ is reflected in 
the spatial extent of the coherent lobes away from the diagonal and 
in the coherence between different wells. Both ratios become 
numerically ill-conditioned in regions where the density is very 
small, such as near the maxima of the lattice potential.

\vspace{0.3cm}
\noindent
\emph{One-body, two-body Information Entropy, and  symmetric Kullback–Leibler divergence}\\

The Shannon information entropy (SIE), is widely used to quantify statistical relaxation~\cite{Berman:2004}. In coordinate space, the SIE of the one-body density is defined as
\begin{equation}
S_x(t) = -\int \mathrm{d}x\, \rho(x,t)\ln[\rho(x,t)],
\end{equation}
%
%
%
%
We formulate an SIE using the two-body diagonal density distribution $\rho^{(2)}(x,x';t) $
\begin{equation}
S_{\rho-x}(t) = -\int \mathrm{d}x\, \mathrm{d}x^{\prime}\,\rho^{(2)}(x,x';t)\ln[\rho^{(2)}(x,x';t)]
\end{equation}
Since $\int\rho\,dx=N$ and $\iint\rho^{(2)}dx\,dx'=N(N-1)$, these quantities are related to the differential entropies $h_1$ and $h_2$ of the unit-normalized distributions $\rho/N$ and $\rho^{(2)}/[N(N-1)]$ by $S_x=N h_1-N\ln N$ and $S_{\rho-x}=N(N-1)\{h_2-\ln[N(N-1)]\}$; as differential entropies, they depend on the choice of the length unit $\bar L$.


To quantify the redistribution of the one-body density following the quench, we use the symmetric Kullback–Leibler (SK) divergence,

\begin{equation}
{\mathrm{SK}}(t)
=
\frac{1}{2}
\int dx\,
\left[
\rho(x,t)
\ln\frac{\rho(x,t)}{\rho(x,0)}
+
\rho(x,0)
\ln\frac{\rho(x,0)}{\rho(x,t)}
\right].
\end{equation}
Here, \(\rho(x,t)\) is the one-body density, normalized such that $\int dx\,\rho(x,t)=N$; with this normalization, SK equals $N$ times the symmetric divergence of the unit-normalized densities.

The SK divergence measures the spatial redistribution of the one-body density relative to its initial configuration, with \({\mathrm{SK}}=0\) when the two density distributions coincide. It therefore provides a complementary density-based diagnostic of the nonequilibrium dynamics.

\vspace{0.3cm}
\noindent
\emph{Many-body Information Entropy}\\

From the time-dependent many-body basis, we compute several measures of many-body information entropy that characterize the redistribution of many-body probability, statistical relaxation, and the degree of irregularity in the ensuing dynamics.

Using the time-dependent expansion coefficients of the state in the MCTDH basis, we define the coefficient Shannon information entropy and the coefficient inverse participation ratio. The squared modulus of each coefficient can be expressed as

\begin{equation}
|C_{\vec{n}}(t)|^2
=
\frac{1}{\prod_{i=1}^{M} n_i!}
\left|
\left\langle 0 \left|
\prod_{i=1}^{M} [\hat b_i(t)]^{n_i}
\right|\Psi(t)\right\rangle
\right|^2.
\end{equation}

%
%
Here $\sum_{k=1}^{M}n_k=N$, and the creation operators stand to the left of the annihilation operators (normal order): since the $N$ annihilation operators map any $N$-particle state onto a multiple of the vacuum, this expression equals $|\langle\vec n;t|\Psi(t)\rangle|^2$. The anti-normally ordered product, as printed in Ref.~\cite{barnali_axel}, equals $\langle\prod_k\binom{\hat n_k+n_k}{n_k}\rangle\ge1$ and is not a probability.\\

The \emph{coefficient Shannon information entropy} is defined as
\begin{equation}
 S_{C}(t) = -\sum_{\vec{n}} | C_{\vec{n}}(t)|^2 \ln |C_{\vec{n}}(t)|^{2},
\end{equation}
which characterizes the distribution of the state $\vert \Psi(t) \rangle$ in the time-dependent Fock space spanned by the permanents of Eq.~(\ref{many_body_wf_2}).\\

A related quantity, the \emph{coefficient inverse participation ratio}, is defined as
\begin{equation}
    I_C(t) = \frac{1} {\sum_{\vec{n}}| C_{\vec{n}}(t)|^4},
\end{equation}
It directly quantifies the effective number of basis states contributing to the many-body dynamics; it is also known as the number of principal components~\cite{barnali_axel}, while some works reserve the term inverse participation ratio for its reciprocal, $\sum_{\vec n}|C_{\vec n}|^4$. It is related to the second R\'enyi entropy of the same distribution by $S_2=\ln I_C$, and $0\le S_2\le S_C\le\ln N_{\mathrm{conf}}$, with $N_{\mathrm{conf}}=48620$ for $N=9$ and $M=10$. A large $I_C$ indicates spreading over many configurations of the chosen basis but does not, by itself, establish ergodicity.

The relevance of $S_C(t)$ and $I_C(t)$ becomes clear in comparison with mean-field theory. In the Gross–Pitaevskii description, only a single configuration contributes, giving $S_C(t) = 0$ and $I_C(t)=1$; such measures are thus inaccessible in mean-field approximations. Consequently, $S_C(t)$ and $I_C(t)$ quantify deviations from mean-field behavior. Small values indicate the many-body state spans few configurations, $|\Psi(t)\rangle$ remains localized in configuration space. Under strong quenches, however, the state can dynamically delocalize across many configurations, increasing entropy. In a finite isolated system, such quantities do not converge to time-independent values, but they may equilibrate in the sense of remaining close to their time averages for most times~\cite{Reimann2008,ShortFarrelly2012}. We therefore apply the same descriptive criterion to both interaction regimes: a measure is regarded as statistically stationary within the simulated window when its late-time values fluctuate around a mean without visible systematic drift, with fluctuations small compared with its overall change after the quench. This criterion refers to the configuration-space measures in the employed representation and does not, by itself, establish equilibration of the many-body state or thermalization.

The coefficients $C_{\vec n}(t)$ refer to the time-dependent orbitals $\psi_k(x;t)$ used in the propagation [Eq.~(\ref{many_body_wf_2})]. A unitary rotation of these orbitals leaves $|\Psi(t)\rangle$ unchanged but changes the coefficients and, in general, $S_C$ and $I_C$; for instance, $\hat a^\dagger\hat b^\dagger|0\rangle$ has $S_C=0$ in the $(a,b)$ basis and $S_C=\ln2$ in the rotated basis $(a\pm b)/\sqrt2$. When the natural occupations are nearly degenerate, as for the initial states considered here, this freedom persists even if natural orbitals are used. $S_C$ is the Shannon entropy of the configuration probabilities in this specified representation; it ignores relative phases and differs from the von Neumann entropy of the pure many-body state, which vanishes. Throughout this work, redistribution of information refers to the spreading of these configuration probabilities. 



\section{Initial states and correlation}

\begin{figure}[!tbh]
    \centering
    \includegraphics[width=1.0\columnwidth]{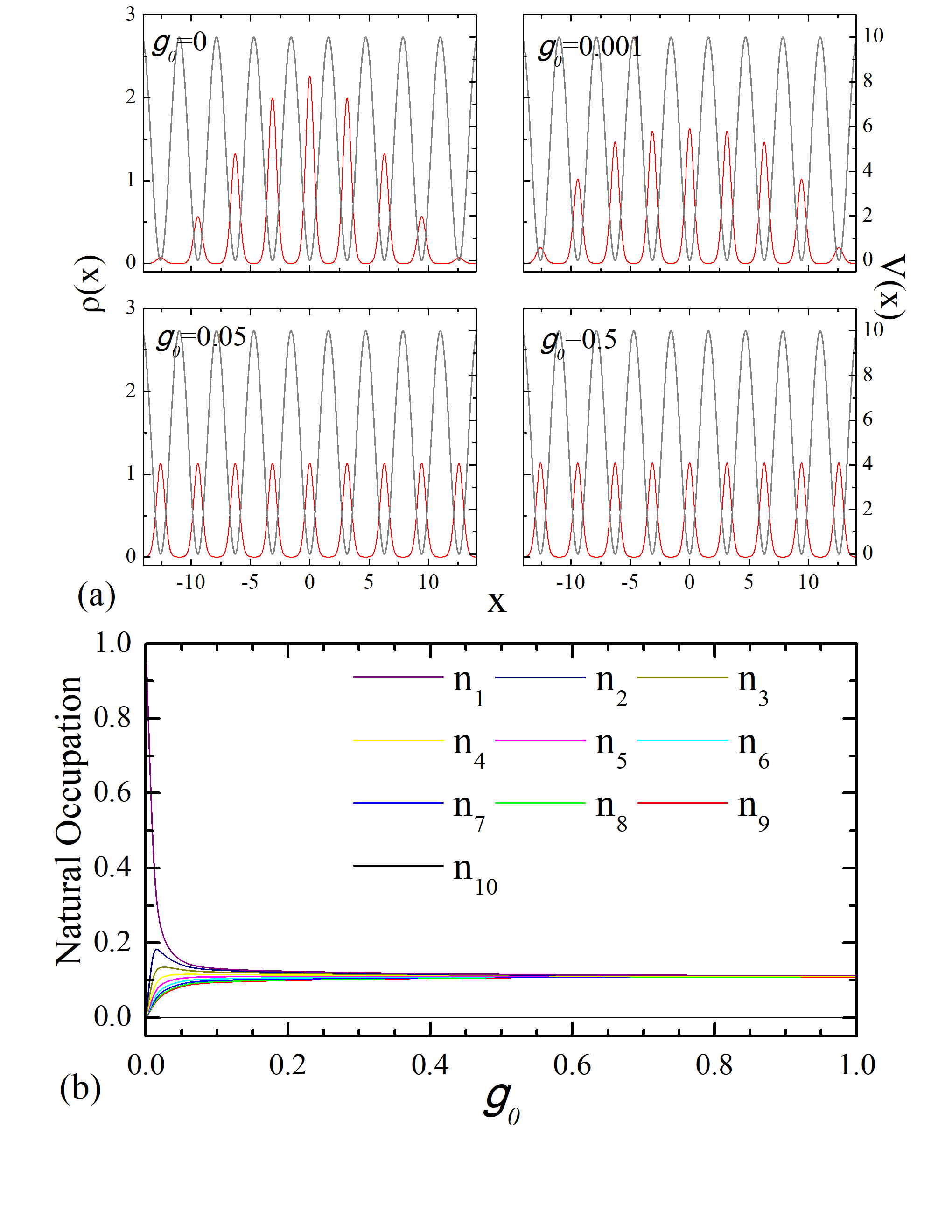}
\caption{(a) One-body density $\rho(x)$ of nine bosons in the primary optical lattice consisting of nine wells, shown for $g_0=0,\,0.001,\,0.05,$ and $0.5$. The primary-lattice potential is shown in the background. The density profiles for $g_0=0.05$ and $0.5$ are essentially identical, indicating the persistence of the Mott-like density profile at strong interaction.
(b) Normalized populations $n_i=\rho_i/N$ of the first ten natural orbitals as a function of $g_0$, with the orbitals ordered by decreasing occupation. The progressive redistribution of population among the natural orbitals with increasing interaction strength characterizes the fragmentation of the initial state.
}
    \label{fig1}
\end{figure}

\begin{figure}[!tbh]
    \centering
    \includegraphics[width=1.0\columnwidth, angle=0]{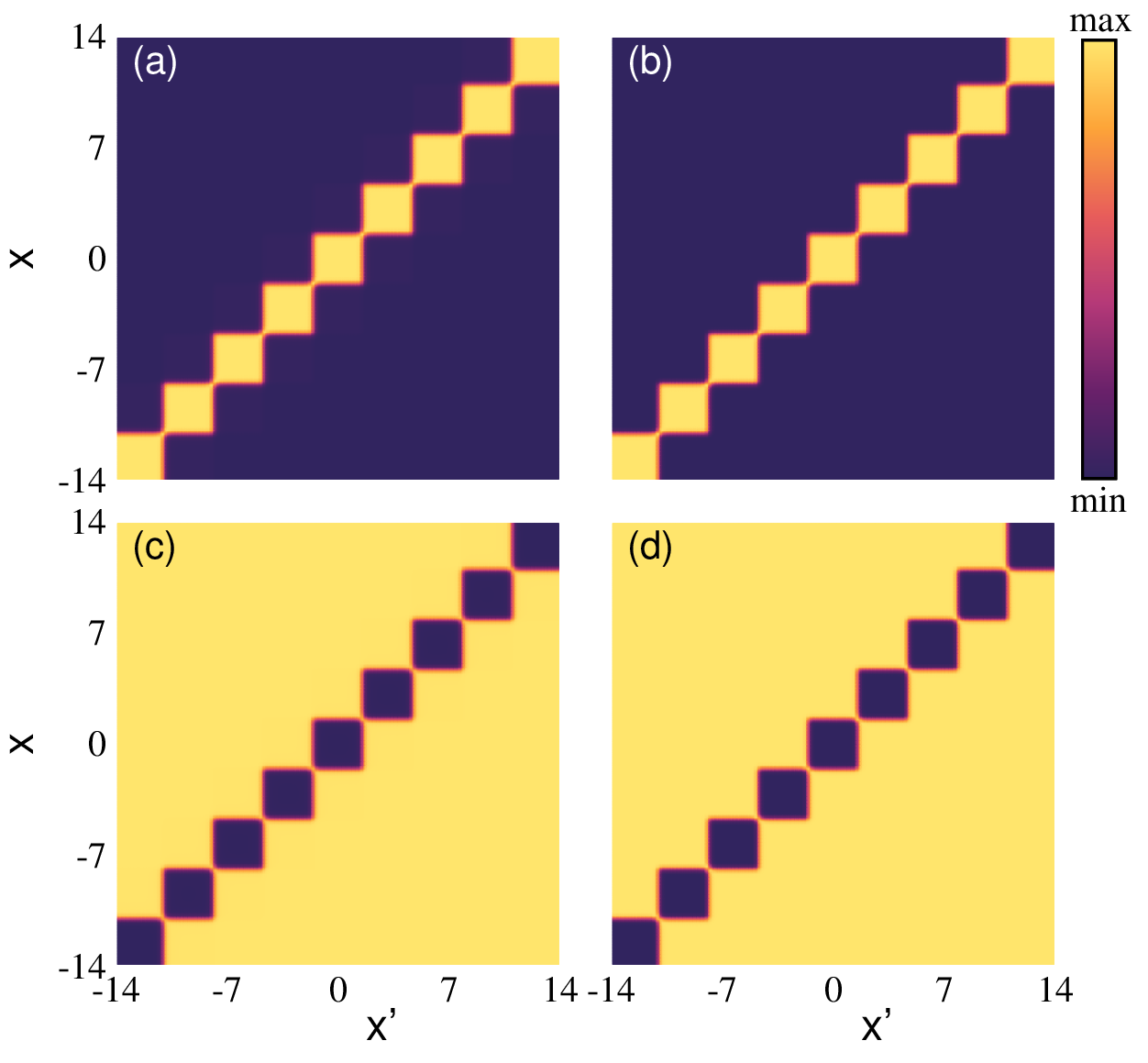}
    \caption{Top panel: Glauber one-body correlation function $|g^{(1)}(x, x')|^{2}$ in the primary lattice for $N = 9$ bosons in $S = 9$ lattice sites with lattice depth $V_p=10$. Panel (a) corresponds to weakly interacting Mott state with $g_0=0.05$ while panel (b) shows the strongly interacting Mott state with $g_0=0.5$.
    Bottom panel: Glauber two-body correlation function for the same parameters as in the top panel. The computations are performed with $M=10$ orbitals. See the text for details.}
    \label{fig2}
\end{figure}

We first examine the ground-state properties of the prequench system in a periodic lattice of depth $V_p=10$ with commensurate filling, $N=S=9$. Interaction strength ranges from noninteracting to partially fragmented Mott state and fully fragmented, strongly correlated Mott state. The one-body density $\rho(x)$ in the primary lattice is shown in Fig.~\ref{fig1}(a) for $g_0 = 0.0,\, 0.001,\, 0.05,$ and $0.5$.

In the periodic lattice, the ground-state configuration results from the competition between kinetic, interaction, and trapping energies. For noninteracting bosons, the density peaks at central sites, forming a superfluid-like state. Weak interactions ($g_0=0.001$) redistribute particles toward outer sites, yielding a fragmented superfluid-like state. At $g_0=0.05$ bosons localize equally across all sites, forming a Mott-insulating state with one atom per site and vanishing density overlap. Further increasing the interaction to $g_0=0.5$ does not change the density profile.

The superfluid–Mott insulator crossover is further characterized by fragmentation. In Fig.~\ref{fig1}(b), we show the occupations of $M=10$ natural orbitals as a function of interaction strength. According to the Penrose–Onsager criterion~\cite{Onsager}, a nonfragmented superfluid occupies only the first orbital, while localized states populate several orbitals, indicating fragmentation. For very weak interactions, the first orbital is macroscopically occupied, and the system is well approximated by a single-orbital mean-field state $\left |N, 0.....\right > $. As $g_0$ increases, the first orbital occupation decreases while others gradually increase. At $g_0=0.05$, when the density is evenly distributed across all sites, several orbitals have significant occupation ($n_1=0.141$, $n_2=0.135$, $n_3=0.126$, $n_4=0.116$, $n_5=0.108$, $n_6=0.102$, $n_7=0.094$, $n_8=0.089$, $n_9=0.087$, with the tenth orbital contributing less than $10^{-6}$), forming a \emph{partially fragmented Mott state}. This label refers to nearly, but not exactly, uniform occupations: the largest occupation, $n_1=0.141$, gives $\mathcal{F}\simeq0.859$ and an occupation entropy $-\sum_j n_j\ln n_j\simeq2.18$, compared with $\mathcal{F}=8/9\simeq0.889$ and an occupation entropy of $\ln 9\simeq2.197$ for uniform occupation.

For $g_0=0.5$, the Mott state becomes \emph{fully fragmented}, with ninefold fragmentation ($n_1 \approx n_2 \approx n_3 \approx n_4 \approx n_5 \approx n_6 \approx n_7 \approx n_8 \approx n_9 \approx 1/9$), reflecting equal eigenvalue per lattice site. The many-body wave function is a nine-fold fragmented state. 


We analyze the first- and second-order coherence functions for the partially ($g_0=0.05$) and fully ($g_0=0.5$) fragmented Mott states, shown in Fig.~\ref{fig2}. The top panels, Figs.~\ref{fig2}(a)–\ref{fig2}(b), depict the first-order coherence function. In both cases, nine bright lobes of finite extent appear around the diagonal, showing that first-order coherence extends over each well (on the diagonal itself, $|g^{(1)}|^2=1$ by definition), while the elements connecting different wells vanish ($|g^{(1)}|^2 \approx 0$), indicating loss of coherence between sites.

The bottom panels, Figs.~\ref{fig2}(c)–\ref{fig2}(d), show the second-order coherence $g^{(2)}$. Between different wells, pair detection is approximately uncorrelated ($g^{(2)} \approx 1$)
and strongly suppressed on the diagonal ($g^{(2)} \approx 0$, antibunching). The nine suppressed diagonal lobes correspond to correlation holes~\cite{Sascha:2010}, consistent with the fact that double occupation of individual wells is strongly suppressed.

The density and correlation functions therefore provide a common signature of Mott localization for both interaction strengths. Within the conventional Bose--Hubbard (BH) description, these two states would be identified as Mott-insulating states with essentially indistinguishable site-localized density and correlation profiles. The single-band BH model nevertheless resolves fragmentation: the eigenvalues of $\Gamma_{ij}=\langle\hat b_i^\dagger\hat b_j\rangle$ are the natural occupations within the lowest band, which become equal ($\Gamma_{ij}=\delta_{ij}$) in the ideal Mott limit, and the two interaction strengths correspond to different values of $U/J$.

The \emph{ab initio} continuum many-body description additionally captures orbital deformation, intrawell structure, and the occupation of higher bands, which are absent from a rigid single-band description. The partially fragmented Mott state at $g_0=0.05$ has a natural occupations that deviate moderately from uniformity (largest fraction $0.141$), whereas at $g_0=0.5$ the population is distributed nearly equally among the relevant natural orbitals. Thus, the natural-occupation spectrum provides a quantitative distinction between the weakly and strongly correlated Mott states that is not visible in their density and correlation profiles; since it changes together with the interaction and quench energy scales, it is not an isolated cause of the dynamics. This distinction becomes important for understanding their markedly different dynamical responses following the subsequent quench into the quasiperiodic lattice.

\section{Nonequilibrium Dynamics at Weak Interaction}

We first examine the nonequilibrium response of the partially fragmented Mott state at \(g_0=0.05\) following the quench into the quasiperiodic lattice. We characterize the dynamics at three complementary levels: conventional observables, spatial correlation functions, and information entropy. This separation allows us to distinguish the evolution of conventional Mott signatures and spatial correlations from the redistribution of information at the one-, two-, and many-body levels.

\subsection{Dynamics of Conventional Observables}

\begin{figure}[!tbh]
    \centering
    \includegraphics[width=1.0\columnwidth]{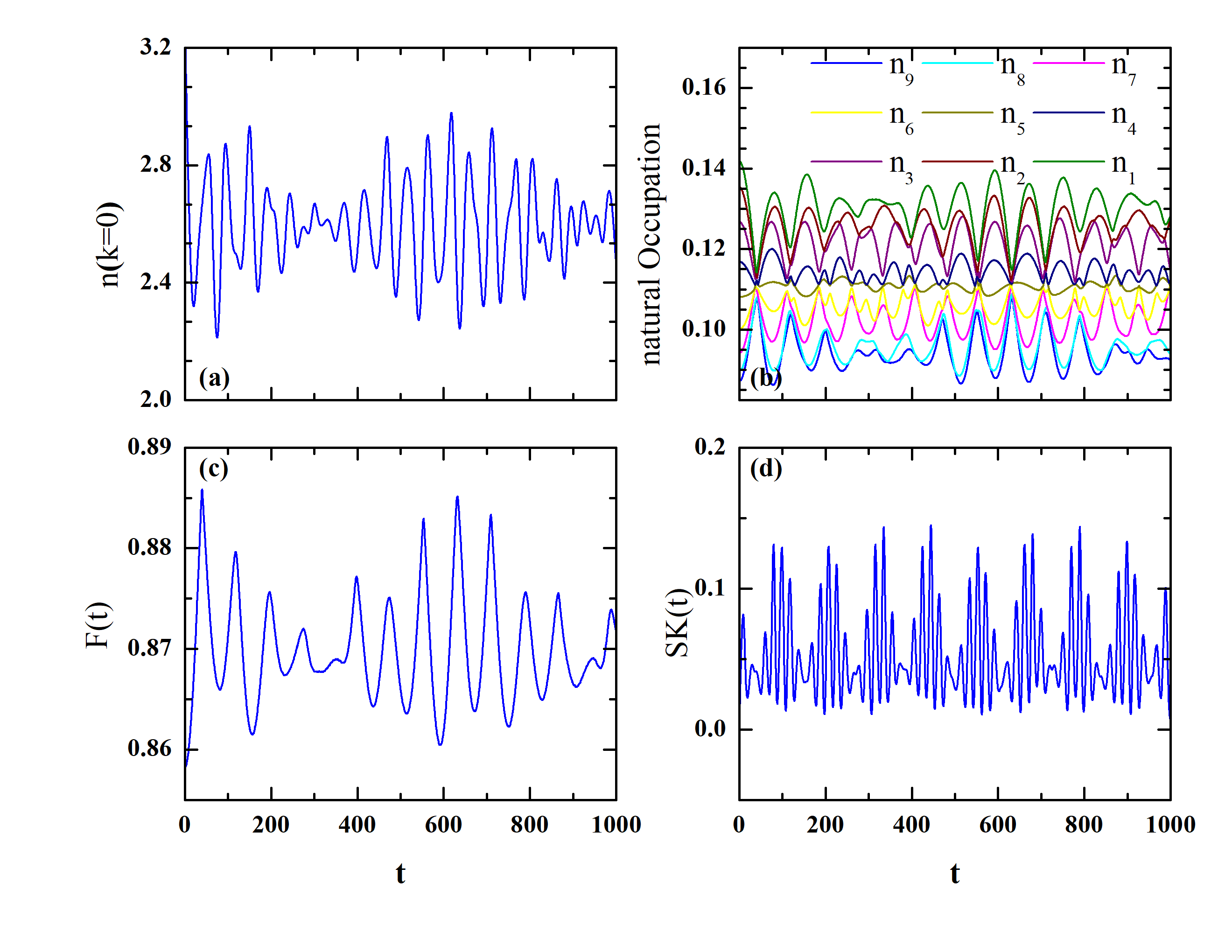}
\caption{Dynamics following a sudden quench in the quasiperiodic lattice for a weakly interacting Mott state with $N=9$ bosons in $S=9$ lattice sites. The parameters are $V_p=10$, $g_0=0.05$, and $V_d=3$, and the simulations are performed using $M=10$ orbitals.
(a) Zero-momentum population $n(k=0,t)$.
(b) Time-dependent normalized populations $n_i(t)$ of the dominant natural orbitals.
(c) Degree of fragmentation $\mathcal{F}(t)=1-n_1(t)$.
(d) Symmetric Kullback--Leibler divergence $\mathrm{SK}(t)$, quantifying the redistribution of the one-body density relative to its initial configuration.
}
    \label{fig3}
\end{figure}

We first examine the nonequilibrium dynamics of the weakly interacting, partially fragmented Mott state at ($g_0=0.05$) following the sudden switch-on of the secondary incommensurate lattice. We consider a representative quench with secondary-lattice depth ($V_d=3.0$). Fig.~\ref{fig3} presents the resulting dynamics through four complementary quantities: the zero-momentum
population $(n(k=0,t))$, the natural-orbital occupations ($n_i(t)$), the degree of fragmentation $(\mathcal{F}(t))$, and the symmetric Kullback--Leibler divergence $({\mathrm{SK}}(t))$.

The zero-momentum population in Fig.~\ref{fig3}(a) exhibits pronounced temporal oscillations following the quench, reflecting the dynamical redistribution of momentum-space population. The oscillatory behavior of \(n(k=0,t)\) reflects repeated redistribution of particles between the zero-momentum mode and finite-momentum components, providing a direct signature of the persistent coherence dynamics induced by the quench.

The dynamical fragmentation provides a direct measure of the redistribution of population among the natural orbitals following the quench. Figure~\ref{fig3}(b) shows the time-dependent occupations of the dominant natural orbitals for \(V_d=3.0\). The nine relevant orbitals exhibit pronounced temporal oscillations, while the tenth orbital remains negligibly populated, \(n_{10}<10^{-6}\), throughout the evolution. The occupations evolve in a correlated manner, with population being transferred among the dominant orbitals during the dynamics. The corresponding degree of fragmentation, \(\mathcal{F}(t)=1-n_1(t)\), shown in Fig.~\ref{fig3}(c), follows these population redistributions and exhibits pronounced temporal variations. Thus, although the initial state is partially fragmented, the quench induces substantial and persistent dynamical redistribution among the natural orbitals.

The symmetric Kullback--Leibler divergence in Fig.~\ref{fig3}(d) also exhibits clear temporal oscillations, but with a broad wave-packet-like structure. Since $\mathrm{SK}$ quantifies the redistribution of the one-body density relative to its initial configuration, this structure reflects the time-dependent spatial redistribution of the density following the quench. Thus, the oscillatory dynamics is consistently visible across the momentum population, natural occupations, fragmentation, and density-based information distance, although each quantity probes a different aspect of the evolving many-body state.

Together, the panels reveal a persistent dynamical competition between the initial Mott-like structure and the quasiperiodic potential. The results shown here correspond to the representative quench \(V_d=3\); the other values of \(V_d\) considered in our simulations exhibit the same qualitative oscillatory behavior (not shown), with the amplitude and timescale of the response depending on the quench strength. The quench transiently modifies the density distributions, but the weakly interacting Mott state does not relax to a stationary configuration within the evolution time; instead, it exhibits persistent oscillatory dynamics. We next examine the corresponding evolution of the spatial correlation functions to determine how the Mott-like correlation structure is affected by the quench.

\subsection{Melting of Mott Correlations}

The persistent oscillatory dynamics observed in the conventional observables does not, by itself, reveal how the underlying Mott correlations evolve following the quench. We therefore examine the first- and second-order correlation functions, \(g^{(1)}(x,x';t)\) and \(g^{(2)}(x,x';t)\), which provide complementary information on the coherence and two-body correlation structure of the evolving state. The suppression of intersite coherence in \(g^{(1)}\) and the diagonal correlation holes in \(g^{(2)}\), associated with suppressed double occupation, are characteristic signatures of the initial Mott state. Their time evolution therefore allows us to directly track the melting and subsequent recovery of the Mott correlation structure induced by the quasiperiodic quench.

\begin{figure}[!tbh]
    \centering
    \includegraphics[width=1.0\columnwidth, angle=0]{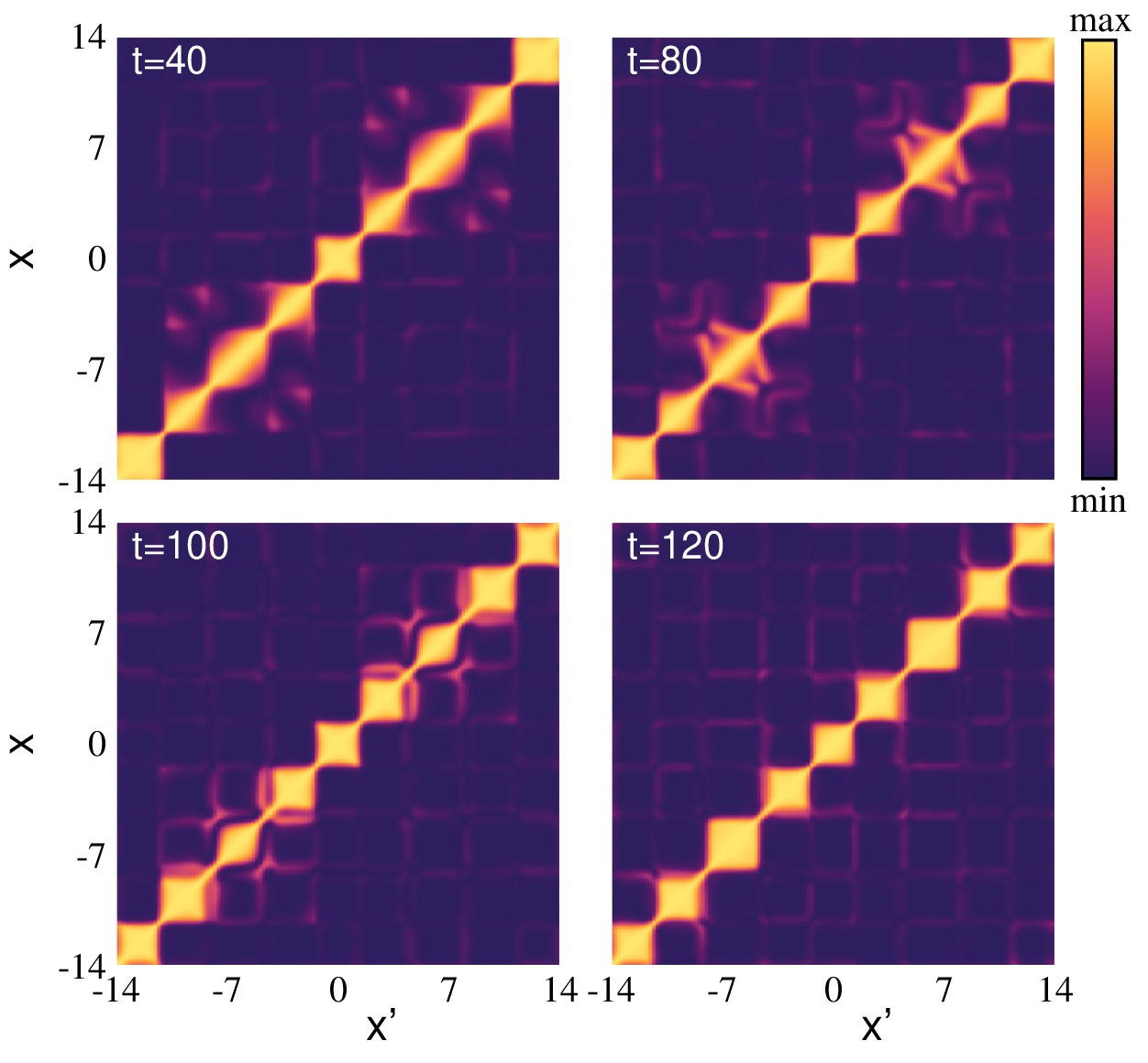}
    \caption{Snapshots of Glauber one-body correlation function $|g^{(1)}(x, x',t)|^{2}$ at selected time instants following a sudden quench in the quasi-periodic lattice. The system consists of $N = 9$ bosons in $S = 9$ lattice sites in the weakly interacting Mott regime. The parameters are: $V_p=10$, $g_0=0.05$, $V_d=3$. The simulations are performed using $M=10$ orbitals. The initially well-defined coherence lobes around the diagonal become distorted following the quench and recover at later times, revealing the collapse and partial revival of the Mott coherence structure. See the text for details.}
    \label{fig4}
\end{figure}

To examine the evolution of coherence following the quench, we first analyze the one-body correlation function \(|g^{(1)}(x,x^{\prime};t)|^{2}\), shown in Fig.~\ref{fig4} at representative times. As discussed in Sec.~V, the initial Mott state in the primary lattice is characterized by nine well-defined lobes around the diagonal, reflecting strong coherence within each lattice site and strongly suppressed coherence between different sites. Following the quench, this characteristic diagonal structure evolves substantially in time. At intermediate times, \(t=40\) and \(t=80\), several lobes become distorted and shrink toward the diagonal, indicating reduced coherence across these wells, accompanied by the appearance of weak finite off-diagonal coherence between neighboring wells. This indicates that the initially localized coherence structure is partially disrupted by the secondary quasiperiodic lattice. At later times, \(t=100\), the lobes begin to recover, and by \(t=120\) the nine bright lobes reemerge. Together with the persistent oscillations of the one-body observables in Fig.~\ref{fig3}, this collapse and partial revival indicates that the weakly interacting Mott state remains dynamically active rather than relaxing to a stationary configuration.

\begin{figure}[!tbh]
    \centering
    \includegraphics[width=1.0\columnwidth, angle=0]{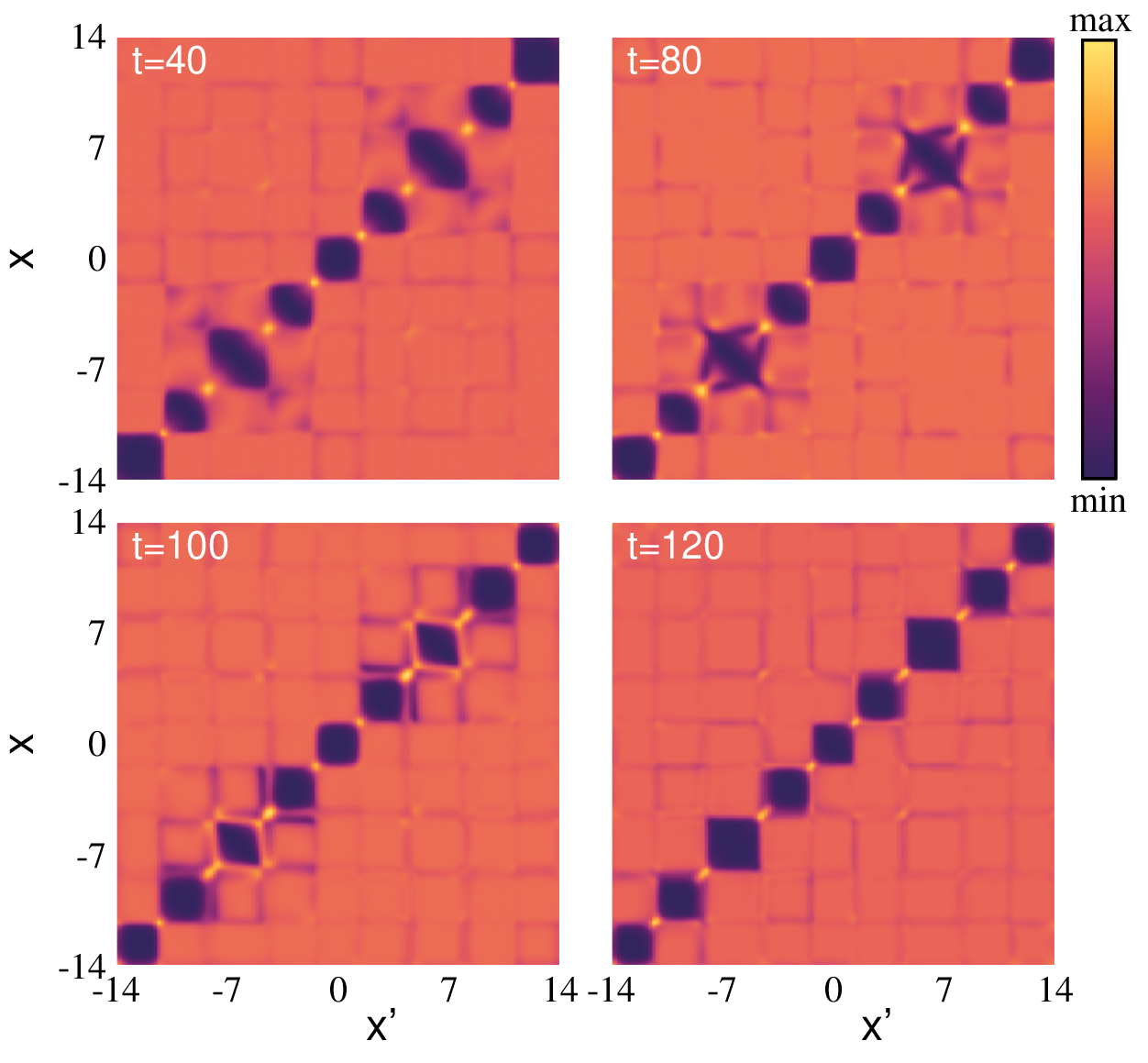}
\caption{Snapshots of the second-order correlation function $g^{(2)}(x,x',t)$ at selected time instants following a sudden quench in the quasi-periodic lattice. The system consists of $N=9$ bosons in $S=9$ lattice sites in the weakly interacting Mott regime. The parameters are $V_p=10$, $g_0=0.05$, and $V_d=3$. The simulations are performed using $M=10$ orbitals. The initially well-defined diagonal correlation holes become distorted during the evolution, indicating a transient melting of the Mott correlation structure, followed by their partial recovery at later times. See the text for details.}
    \label{fig5}
\end{figure}

To directly examine the evolution of the Mott correlations, we next consider the second-order correlation function \(g^{(2)}(x,x^{\prime};t)\) at the same representative times shown in Fig.~\ref{fig5}. In the initial Mott state, \(g^{(2)}\) exhibits nine pronounced diagonal correlation holes, corresponding to the suppression of finding two bosons at the same position and, in the lattice picture, to strongly suppressed double occupation of each site. Following the quench, these diagonal holes progressively become distorted and less sharply defined. At intermediate times, \(t=40\) and \(t=80\), the initially well-localized correlation holes broaden and lose their regular diagonal structure, indicating a partial reduction of the initial Mott correlations. At later times, \(t=100\) and \(t=120\), the diagonal structure becomes more pronounced again, consistent with the partial recovery of the correlation-hole pattern. Thus, the second-order correlations exhibit a collapse-and-revival dynamics that closely accompanies the evolution of the one-body coherence function. The Mott correlations therefore undergo substantial temporal melting and partial recovery at the times shown rather than relaxing monotonically toward a stationary state.

\subsection{Information-Theoretic Diagnostics of Relaxation}

The evolution of the conventional observables and correlation functions discussed above demonstrates substantial nonequilibrium dynamics, but these quantities alone do not establish whether the many-body state relaxes. To address this question, we consider information-entropy measures at different levels of description. The one- and two-body information entropies characterize the redistribution of information contained in reduced one- and two-body descriptions and are therefore sensitive to the evolution of coherence and pair correlations. However, their dynamics alone are insufficient to determine whether the full many-body wavefunction has relaxed.

A complementary diagnostic is provided by the many-body measures \(S_C(t)\) and \(I_C(t)\). In contrast to the one- and two-body entropies, these quantities retain information about the structure of the full many-body state and are sensitive to the redistribution of the wavefunction among many-body configurations. In particular, the coefficient entropy \(S_C(t)\), defined from the time-dependent expansion coefficients of the many-body state, quantifies the spreading of the wavefunction in configuration space. Small values of \(S_C(t)\) correspond to a relatively localized distribution in configuration space, whereas an increase in \(S_C(t)\) indicates that an increasing number of many-body configurations contribute significantly to the evolving state. The combined behavior of \(S_C(t)\) and \(I_C(t)\) therefore provides a more sensitive probe of many-body redistribution in the employed representation than the corresponding reduced one- and two-body measures.

\begin{figure}[!tbh]
    \centering
         \includegraphics[width=\columnwidth]{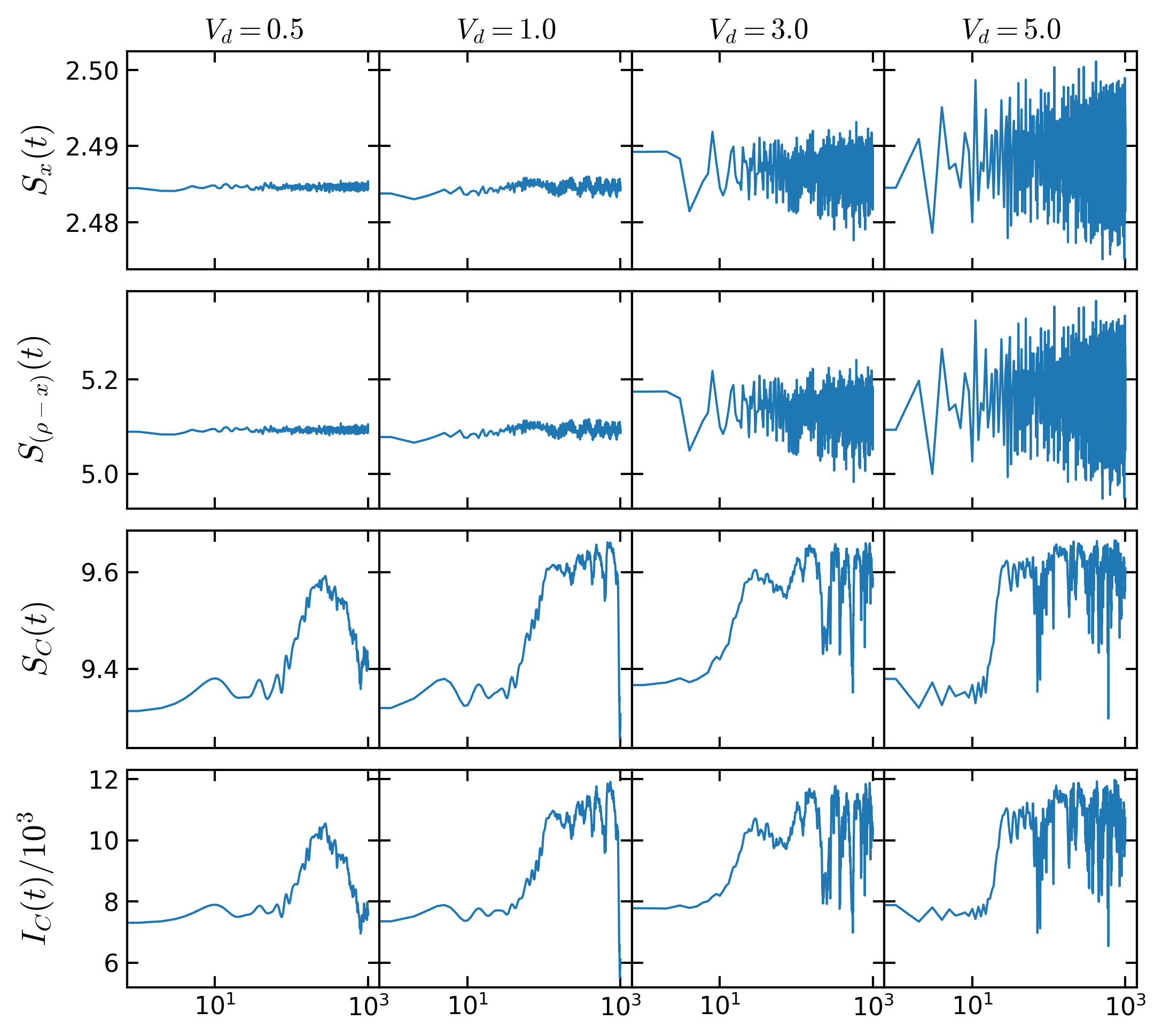}
        \caption{Time evolution of the one-body entropy \(S_x(t)\), two-body entropy \(S_{\rho-x}(t)\), many-body coefficient entropy \(S_C(t)\), and inverse participation ratio \(I_C(t)\) following a sudden quench in the quasiperiodic lattice. The system consists of \(N=9\) bosons in \(S=9\) lattice sites in the weakly interacting Mott regime. The parameters are \(V_p=10\), \(g_0=0.05\), and \(V_d=0.5,\,1,\,3\), and \(5\) (columns, from left to right); the rows show \(S_x\), \(S_{\rho-x}\), \(S_C\), and \(I_C/10^3\) on a logarithmic time axis. The simulations are performed using \(M=10\) orbitals. See the text for details.}
    \label{fig6}
\end{figure}

Fig.~\ref{fig6} summarizes the time evolution of the one-body entropy \(S_x(t)\), the two-body entropy \(S_{\rho-x}(t)\), the many-body coefficient entropy \(S_C(t)\), and the inverse participation ratio, \(I_C(t)\) for four representative quench strengths, \(V_d=0.5,\,1,\,3,\) and \(5\). A clear distinction emerges between the reduced and many-body measures. The one- and two-body entropies exhibit qualitatively similar temporal behavior for all quench strengths. For the weak quench \(V_d=0.5\), both \(S_x(t)\) and \(S_{\rho-x}(t)\) show only a small transient response following the quench and subsequently evolve with comparatively featureless dynamics. With increasing \(V_d\), the initial transient becomes progressively more pronounced, followed by stronger oscillations about a well-defined mean value. Thus, although the amplitude of the temporal response increases with the quench strength, the one- and two-body entropies retain essentially the same qualitative character and do not, by themselves, provide an unambiguous signature of many-body relaxation.

The behavior of the many-body quantities \(S_C(t)\) and \(I_C(t)\) is qualitatively different. At weak quench strength, \(V_d=0.5\), the many-body dynamics remain relatively restricted, with \(S_C(t)\) and \(I_C(t)\) showing only limited redistribution following the initial transient. As the quench strength is increased to \(V_d=1\) and \(3\), the many-body measures exhibit progressively stronger temporal fluctuations and a substantially larger redistribution of information in configuration space. The oscillations of \(S_C(t)\) and \(I_C(t)\) become increasingly pronounced, revealing a many-body response that is not resolved by the comparatively featureless one- and two-body entropies.

For the strongest quench, \(V_d=5\), the many-body measures display the most pronounced redistribution and temporal fluctuations. Importantly, however, the dynamics remain fluctuating rather than converging to a strictly time-independent value. Thus, the strong quench enhances the spreading of the many-body wavefunction over configuration space. For \(V_d=3\) and \(5\), the typical late-time values (\(S_C\approx9.6\), \(I_C\approx1.1\times10^4\)) are comparable to those reached at strong interaction (Fig.~\ref{fig10}), but they are interrupted by large, intermittent downward excursions (down to \(S_C\approx9.3\)--\(9.4\)); for \(V_d\le1\), the measures evolve nonmonotonically, rising and then decreasing within the simulated window.

The comparison therefore reveals a clear hierarchy in the information dynamics. The one- and two-body entropies, \(S_x(t)\) and \(S_{\rho-x}(t)\), primarily reflect the redistribution of reduced coherence and pair-correlation information and exhibit a common, relatively featureless transient-to-oscillatory response. In contrast, \(S_C(t)\) and \(I_C(t)\) retain distinct signatures of the spreading and redistribution of the full many-body wavefunction. Their stronger and increasingly structured response with \(V_d\) indicates that the quench drives substantial many-body reorganization even when the reduced information measures provide little indication of it. Consequently, the many-body information measures offer a more sensitive quantitative diagnostic of the nonequilibrium relaxation dynamics, in the employed representation, than the one- and two-body entropies alone.

Taken together, Figs.~\ref{fig3}–\ref{fig6} provide a consistent picture of the nonequilibrium dynamics of the weakly correlated Mott insulator in the quasiperiodic lattice. Across the conventional observables, correlation functions, and information-entropy measures, the dynamics are dominated by a persistent competition between the correlations inherited from the initial Mott state and those induced by the secondary lattice. This competition gives rise to pronounced and recurrent oscillations in phase coherence, fragmentation, and two-body correlations, accompanied by persistent fluctuations in the many-body information measures.

Importantly, according to the criterion of Sec.~IV, the entropy measures do not exhibit a clear convergence toward a statistically stationary regime within the simulated time window: although their typical late-time values for \(V_d=3\) and \(5\) are comparable to those at strong interaction, the large intermittent excursions and the nonmonotonic evolution for weaker quenches prevent the identification of a stable plateau. Thus, despite substantial nonequilibrium dynamics and transient redistribution of information, the weakly interacting fragmented Mott state does not reach a statistically stationary regime of the many-body entropies within the simulated window. The persistent and structured temporal dynamics reflect a recurrent, coherent reorganization of the state rather than a monotonic approach to a plateau; longer simulations and quantitative stationarity tests are required to establish whether this behavior persists or signals nonergodic dynamics.


\section{Nonequilibrium Dynamics at Strong Interaction}

We next examine the nonequilibrium dynamics of the strongly interacting Mott state at \(g_0=0.5\) following the same sudden quench into the quasiperiodic lattice. In contrast to the weakly interacting case, the initial state is fully fragmented, with nearly uniform natural occupations, and possesses well-developed Mott correlations. We again characterize the dynamics at three complementary levels: conventional observables, spatial correlation functions, and many-body information measures. This comparison allows us to determine whether the persistence of the Mott correlation structure is accompanied by a suppression of many-body information redistribution or whether these two aspects of the dynamics can evolve independently.

\subsection{Dynamics of Conventional Observables}

\begin{figure}[!tbh]
    \centering
    \includegraphics[width=\columnwidth]{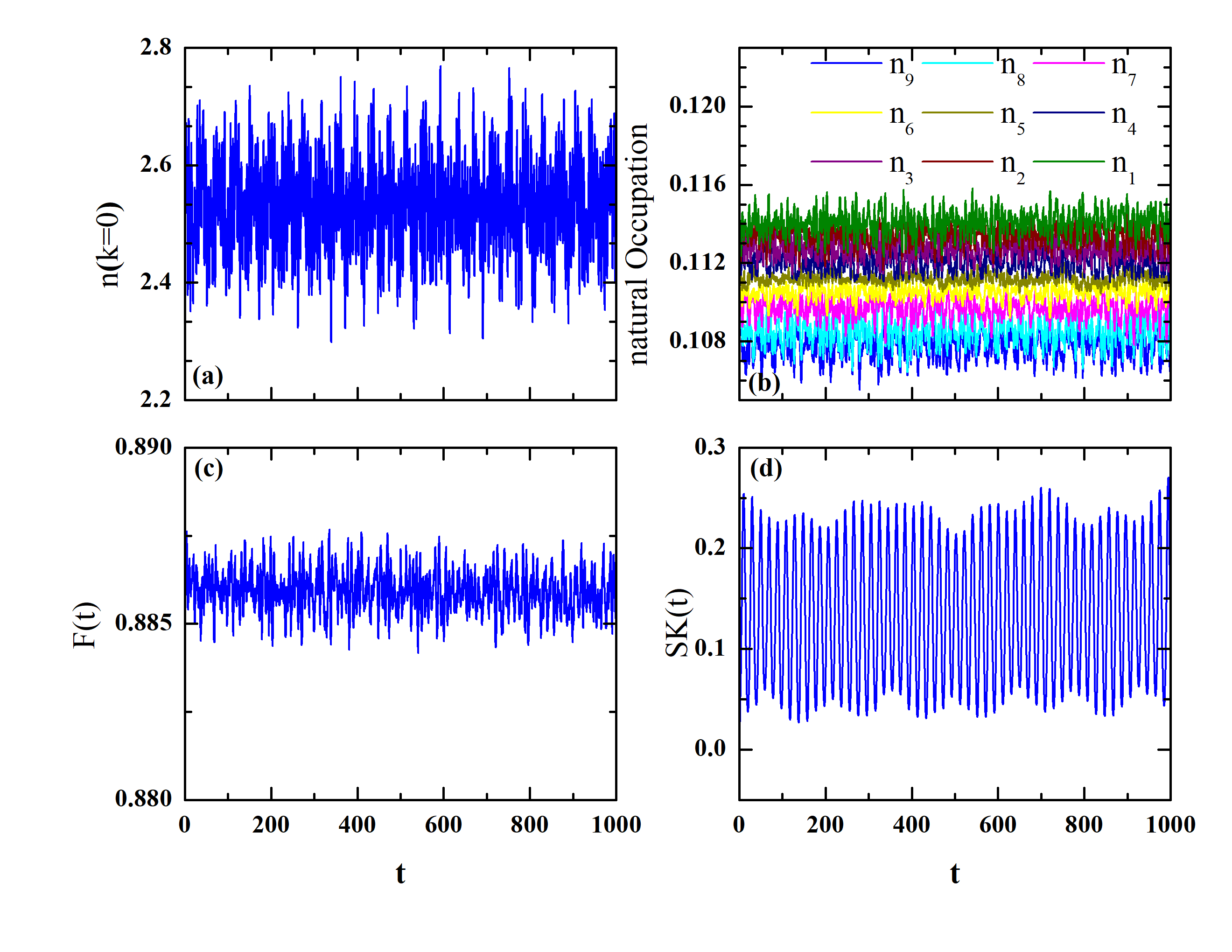}
     \caption{Dynamics following a sudden quench in the quasiperiodic lattice for a strongly interacting Mott state with $N=9$ bosons in $S=9$ lattice sites. The parameters are $V_p=10$, $g_0=0.5$, and $V_d=3$, and the simulations are performed using $M=10$ orbitals.
(a) Zero-momentum population $n(k=0,t)$.
(b) Time-dependent normalized populations $n_i(t)$ of the dominant natural orbitals.
(c) Degree of fragmentation $\mathcal{F}(t)$.
(d) Symmetric Kullback--Leibler divergence $\mathrm{SK}(t)$, quantifying the redistribution of the one-body density relative to its initial configuration.
}
    \label{fig7}
\end{figure}

We first examine the time evolution of the conventional one-body observables following the quench. Fig.~\ref{fig7} shows the dynamics of the zero-momentum occupation, natural-orbital occupations, fragmentation, and the symmetric Kullback--Leibler divergence for a representative quench strength, $V_d=3$.

Fig.~\ref{fig7}(a) shows the time evolution of the zero-momentum peak, \(n(k\!\to\!0,t)\), which exhibits only rapid, irregular fluctuations of smaller amplitude than at weak interaction around its mean value, \(\langle n(k\!\to\!0,t)\rangle\approx2.55\). The weak temporal variation indicates that the quench produces only a limited redistribution of the momentum-space population in the strongly interacting regime.

The corresponding natural occupations, shown in Fig.~\ref{fig7}(b), show that the initially fully fragmented Mott state remains essentially fully fragmented throughout the evolution. The occupations exhibit only weak temporal variations, between about $0.106$ and $0.116$, indicating that the strong interaction maintains the underlying fragmented character of the state and prevents any substantial recovery of single-orbital coherence.

Fig.~\ref{fig7}(c) shows the time evolution of the degree of fragmentation, \(\mathcal F(t)=1-n_1(t)\). The fragmentation remains nearly constant throughout the evolution, with only small temporal fluctuations around its initial value. This indicates that the strongly interacting Mott state retains its highly fragmented character following the quench, with no substantial transfer of population toward a dominant natural orbital. Thus, despite the nonequilibrium evolution induced by the quasiperiodic lattice, the strong interaction robustly preserves the fragmented Mott structure.

Fig.~\ref{fig7}(d) shows the time evolution of the symmetric Kullback--Leibler divergence $\mathrm{SK}(t)$. From the onset of the quench, \(\mathrm{SK}(t)\) displays fast, regular oscillations with a nearly constant envelope, between about $0.03$ and $0.26$. Their amplitude is not smaller than at weak interaction [Fig.~\ref{fig3}(d)], but they lack the slower, irregular modulation observed there, indicating a bounded, periodic redistribution of the one-body density rather than a progressive departure from its initial profile. Thus, the strong interaction preserves the one-body density profile on average, even though, as shown by the many-body measures, the underlying many-body wavefunction can continue to evolve dynamically.

Taken together, Figs.~\ref{fig7}(a)–\ref{fig7}(d) show that the strong interaction strongly constrains the response of the Mott state to the quasiperiodic quench. The zero-momentum occupation and degree of fragmentation exhibit only weak temporal variations, while the symmetric Kullback--Leibler divergence oscillates regularly within a bounded band from the onset of the evolution. These results indicate that the quench produces only limited and regular changes in the conventional one-body observables, in marked contrast to the pronounced and irregular oscillatory dynamics observed in the weakly interacting regime. The strong interaction therefore preserves the dominant low-order properties of the initial Mott state, despite the presence of the quasiperiodic perturbation.

\subsection{Persistence of Mott Correlations}

\begin{figure}[!tbh]
    \centering
    \includegraphics[width=\columnwidth, angle=0]{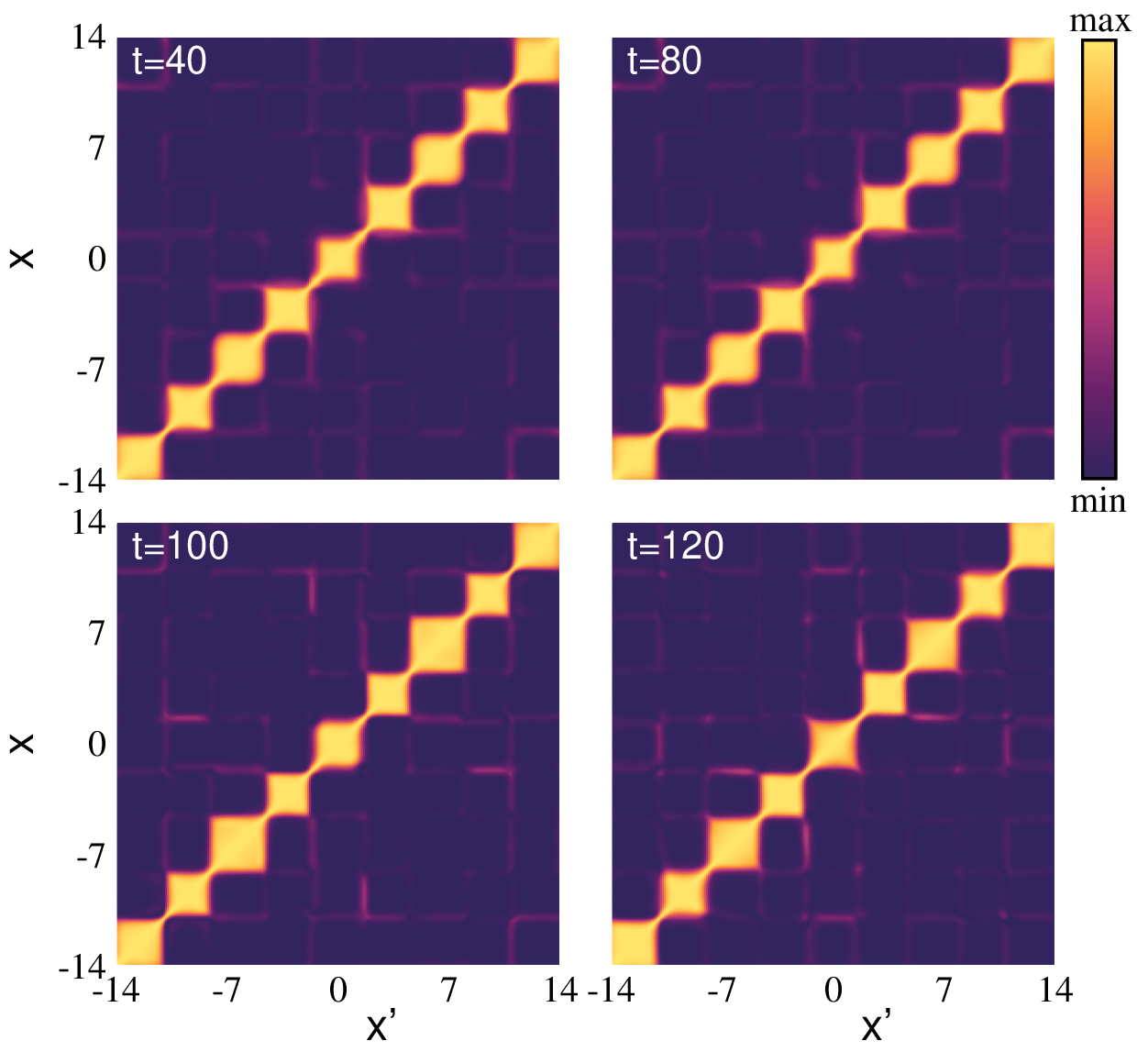}
    \caption{
Snapshots of Glauber one-body correlation function $|g^{(1)}(x, x';t)|^{2}$ at selected time instants following a sudden quench in the quasi-periodic lattice. The system consists of $N = 9$ bosons in $S = 9$ lattice sites in the strongly interacting Mott regime. The parameters are: $V_p=10$, $g_0=0.5$, $V_d=3$. The simulations are performed using $M=10$ orbitals. See the text for details.
    }
    \label{fig8}
\end{figure}

\begin{figure}[!tbh]
    \centering
    \includegraphics[width=1.0\columnwidth, angle=0]{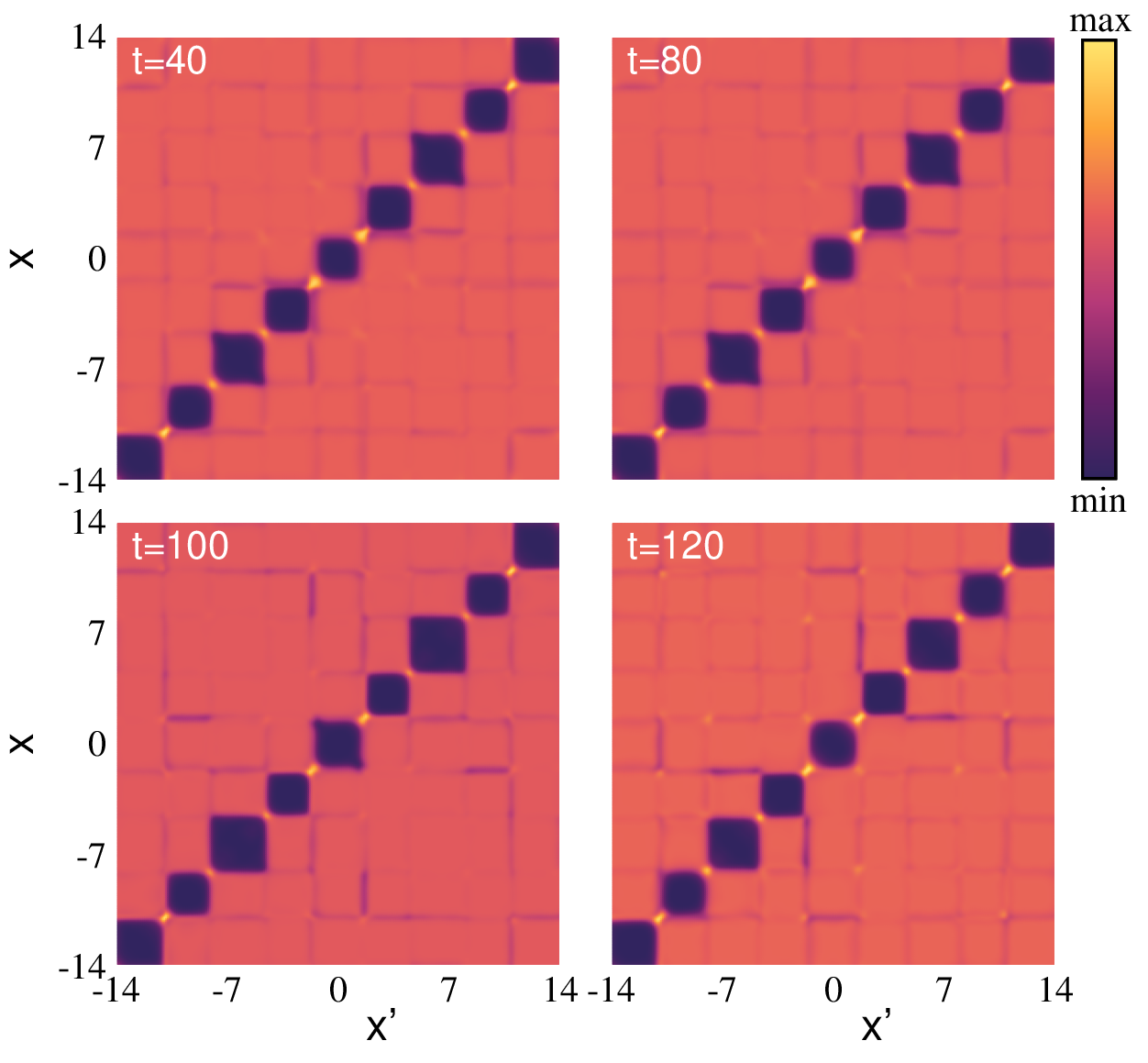}
\caption{Snapshots of the second-order correlation function $g^{(2)}(x,x',t)$ at selected time instants following a sudden quench in the quasi-periodic lattice. The system consists of $N=9$ bosons in $S=9$ lattice sites in the strongly interacting Mott regime. The parameters are $V_p=10$, $g_0=0.5$, and $V_d=3$. The simulations are performed using $M=10$ orbitals. See the text for details.}
    \label{fig9}
\end{figure}

To further assess the robustness of the Mott correlations, Figs.~\ref{fig8} and \ref{fig9} show the one- and two-body correlation functions, \(\lvert g^{(1)}(x,x';t)\rvert^{2}\) and \(g^{(2)}(x,x';t)\), respectively, at selected times. In contrast to the pronounced temporal deformation of the correlation structure observed in the weakly interacting regime, the strongly interacting state exhibits a persistence of its initial Mott character at the times shown, with partial distortions. In \(\lvert g^{(1)}(x,x';t)\rvert^{2}\), the off-diagonal elements remain strongly suppressed, indicating negligible phase coherence between different lattice sites, while the lobes around the diagonal remain confined to individual wells. The second-order correlation function \(g^{(2)}(x,x';t)\) provides complementary evidence: the pronounced diagonal correlation holes associated with suppressed double occupation remain clearly visible at the times shown, with only partial temporal deformation.

Together, the persistence of suppressed intersite coherence in \(g^{(1)}(x,x';t)\) and the diagonal correlation holes in \(g^{(2)}(x,x';t)\), combined with the nearly constant degree of fragmentation in Fig.~\ref{fig7}(c), indicate that the strong interaction robustly preserves the local Mott correlation structure despite the quasiperiodic quench. Thus, the conventional correlation signatures remain stable, up to partial distortions, even though, as shown below, the full many-body wavefunction can undergo substantial dynamical redistribution.

\subsection{Information-Theoretic Diagnostics of Relaxation}

\begin{figure}[!tbh]
    \centering
         \includegraphics[width=\columnwidth]{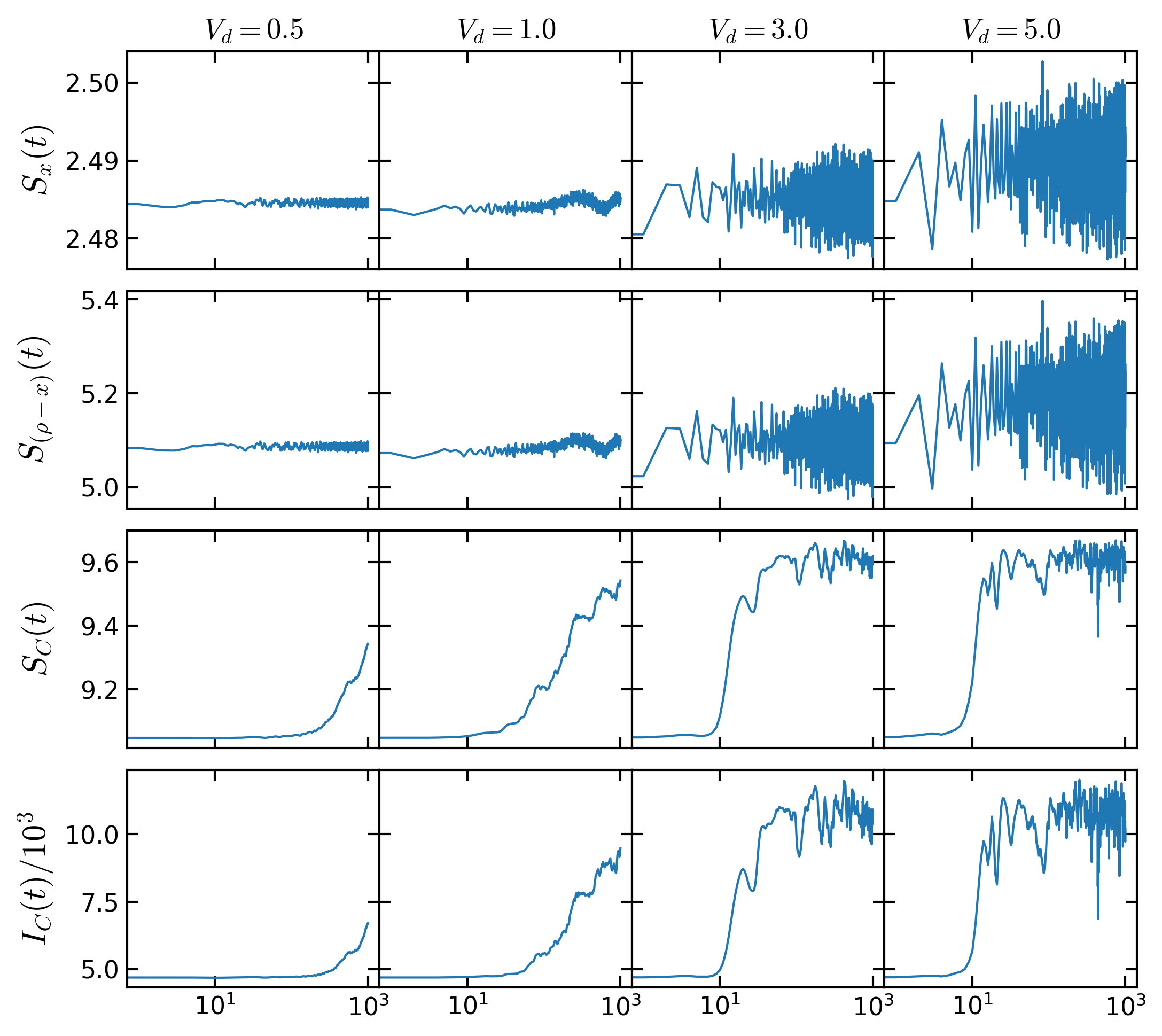}
    \caption{Time evolution of the one-body entropy \(S_x(t)\), two-body entropy \(S_{\rho-x}(t)\), many-body coefficient entropy \(S_C(t)\), and inverse participation ratio \(I_C(t)\) following a sudden quench in the quasiperiodic lattice. The system consists of \(N=9\) bosons in \(S=9\) lattice sites in the strongly interacting Mott regime. The parameters are \(V_p=10\), \(g_0=0.5\), and \(V_d=0.5,\,1,\,3\), and \(5\) (columns, from left to right); the rows show \(S_x\), \(S_{\rho-x}\), \(S_C\), and \(I_C/10^3\) on a logarithmic time axis. The simulations are performed using \(M=10\) orbitals. See the text for details.}
    \label{fig10}
\end{figure}

The persistence of the Mott correlation structure does not, by itself, determine whether the full many-body state undergoes relaxation. We therefore examine the information-entropy measures at different levels of description. Fig.~\ref{fig10} summarizes the time evolution of the one-body entropy \(S_x(t)\), two-body entropy \(S_{\rho-x}(t)\), many-body coefficient entropy \(S_C(t)\), and inverse participation ratio \(I_C(t)\) for four representative quench strengths, \(V_d=0.5,\,1,\,3,\) and \(5\).

The one- and two-body entropy measures, \(S_x(t)\) and \(S_{\rho-x}(t)\), exhibit relatively modest temporal variations for all quench strengths. For \(V_d=0.5\), their dynamics remain comparatively weak and regular, while increasing \(V_d\) produces stronger fluctuations. Nevertheless, their qualitative behavior remains similar across the different quenches and does not by itself reveal the relaxation of the full many-body wavefunction.

In contrast, the many-body measures \(S_C(t)\) and \(I_C(t)\) exhibit a pronounced dependence on the quench strength. For weak quenches, the redistribution of the many-body wavefunction remains strongly restricted and is still growing at the end of the simulated window, whereas for large \(V_d\) the dynamics develop a qualitative multistage relaxation process. In particular, the evolution can be separated into three temporal regimes, as also reported for this interaction strength in Ref.~\cite{Marino2026preprint}; we do not extract crossover times. At short times, the many-body measures exhibit a rapid initial response following the quench, reflecting the onset of redistribution among the many-body configurations. This is followed by an intermediate regime in which the information measures undergo pronounced temporal evolution and the wavefunction progressively spreads over configuration space. At longer times, the dynamics enter a late-time regime in which \(S_C\approx9.6\) and \(I_C\approx1.1\times10^4\) fluctuate around nearly constant mean values, with fluctuations much smaller than their increase after the quench.

The emergence of these three temporal regimes becomes increasingly pronounced as the quench strength is increased. Thus, the secondary lattice controls not only the overall magnitude of the many-body information redistribution but also the temporal structure of the relaxation process. Importantly, this relaxation occurs while the degree of fragmentation and the spatial correlation structure remain largely preserved. The many-body wavefunction can therefore undergo substantial redistribution in configuration space without destroying the local Mott correlations.

These results indicate that statistical relaxation of the many-body entropies in the strongly interacting Mott state is not synonymous with the melting of its Mott correlations. Instead, the quench induces a hierarchical redistribution of many-body information, with the three-stage dynamics becoming clearly resolved for sufficiently strong \(V_d\), while the fragmented Mott structure remains preserved, up to partial distortions of the correlation maps, throughout the evolution.
The late-time values are consistent with those reported in Ref.~\cite{Marino2026preprint}, $S_C\approx9.6$ and $S_2\approx9.23$ (note that $I_C=e^{S_2}$). The Gaussian Orthogonal Ensemble-based dimension estimates $e^{S_C}/0.48\approx3.1\times10^4$ and $3e^{S_2}\approx3.1\times10^4$ agree with each other but exceed the nominal nine-mode dimension, $\binom{17}{9}=24310$, by about $26\%$; they should be regarded as benchmarks rather than as evidence of ergodicity.

The coexistence of suppressed transport with continuing many-body evolution has important precedents in many-body localization~\cite{Abanin}: interaction-induced dephasing can produce slow entanglement growth while transport remains suppressed~\cite{Bardarson2012,Serbyn2013}. Lukin \emph{et al.}~\cite{Lukin2019} prepared a Mott state with one boson per site in a quasiperiodic potential and observed distinct time scales for the number entropy and the configurational entanglement entropy. That configurational entanglement entropy is not the Shannon entropy of the global expansion coefficients used here, and the experiment quenched the tunneling rather than switching on a secondary lattice. Dephasing in a localized regime is nevertheless a relevant alternative interpretation, which the present diagnostics neither identify nor exclude.


\section{Conclusion}

Understanding how isolated quantum many-body systems redistribute information and approach relaxation following a quantum quench remains a central problem in nonequilibrium quantum physics. In particular, the relationship between the evolution of conventional correlation signatures and the redistribution of information within the full many-body wavefunction is not generally captured by reduced one- or two-body observables. Here, we have addressed this question for interacting bosons in a commensurate Mott state subjected to a sudden quench from a primary optical lattice to a quasiperiodic lattice.

Using an ab initio many-body description, we have compared weakly and strongly interacting Mott states that exhibit similar one-body density profiles and characteristic Mott-like correlation structures in the initial periodic lattice, but differ moderately in their natural-occupation spectra. This provides a controlled setting in which the role of interaction strength in the subsequent nonequilibrium dynamics can be compared. We characterize the evolution through conventional observables, one- and two-body correlation functions, and many-body information measures, including the coefficient entropy and inverse participation ratio.

For the weakly interacting Mott state, the quasiperiodic quench produces pronounced and recurrent nonequilibrium dynamics. The one-body observables and correlation functions exhibit coherent oscillations, while the characteristic Mott correlation structure undergoes a recurrent melting and partial recovery. The corresponding information measures show large, intermittent temporal fluctuations without sustained growth toward a stationary maximum or clear convergence and no stable plateau within the simulated time window, although their typical late-time values for the strongest quenches are comparable to those at strong interaction. Thus, the weakly interacting state does not reach a statistically stationary regime of the many-body entropies within the simulated window, despite exhibiting substantial dynamical redistribution of coherence, fragmentation, and correlations.

The strongly interacting Mott state displays a qualitatively different response. The conventional one-body observables remain close to their initial values on average, with small or regular oscillations, the degree of fragmentation changes only weakly, and the one- and two-body correlation functions retain their characteristic Mott structure, up to partial distortions, at the times shown. In particular, the suppression of intersite coherence and the diagonal correlation holes remain robust under the quasiperiodic quench. Nevertheless, the many-body information measures reveal substantial redistribution of the full many-body wavefunction in configuration space. For sufficiently strong quenches, this redistribution develops a qualitative three-stage temporal structure, consisting of an initial response, an intermediate regime of pronounced many-body spreading, and a late-time plateau with comparatively small fluctuations.

These results indicate an important separation between local correlation dynamics and many-body information relaxation. The persistence of Mott correlations does not imply that the full many-body state remains dynamically frozen. Conversely, substantial redistribution of the many-body wavefunction does not require the destruction of the underlying Mott structure. The interaction strength and quench amplitude therefore provide jointly control the extent and timescale of many-body information redistribution, while the strong interaction can protect the local Mott character against the quasiperiodic perturbation.

Our results indicate that information entropy and configuration-space participation measures provide a sensitive complement to conventional observables for characterizing nonequilibrium many-body dynamics. In particular, they reveal dynamical processes that remain largely hidden in one-body densities, fragmentation, and spatial correlation functions. The observed three-stage relaxation at strong quasiperiodic quenches further provides a qualitative temporal picture of how many-body information is redistributed following the quench.

The dynamics considered here are accessible in principle through time-resolved quench and correlation measurements. Several directions remain open, including a systematic investigation of the crossover between restricted and multistage relaxation, the dependence of the relaxation dynamics on filling and system size, and the characterization of the weak-quench regime over longer times. Extending the analysis to incommensurate fillings, where mobile particles coexist with a commensurate Mott background, may further reveal how localized correlations and mobile components compete during nonequilibrium evolution.

The present conclusions are limited by the dependence of the coefficient-based measures on the orbital representation, by the fixed relative phase and parity sector of the bichromatic potential, and by the small system size. Useful controls include a projection onto a fixed prequench basis of localized orbitals, with the captured norm reported, the fidelity for both interaction strengths, a calibrated Bose--Hubbard comparison, intermediate interaction strengths, and site-resolved number fluctuations. These controls would help to discriminate dephasing in a localized regime from other mechanisms and to establish the physical content of the configuration-space spreading.

\section*{Acknowledgments}
N.~D.~Chavda acknowledges financial support from University Research
Project Grant~No.DR/Dir./26-27/17/Sr~No-5. 
\appendix
\section{System Parameters}



\begin{table}[tbh]
    \centering
    \begin{tabular}{ || c | c || }
        \hline \hline
            Quantity & MCTDH-X units  \\
        \hline \hline
            unit of length & $\bar{L} = 1/k_p \simeq 84.7$~nm \\
        \hline
            unit of energy & $\bar{E} = \frac{\hbar^2}{ m \bar{L}^2} =2E_r$ \\
        \hline
            unit of time & $\bar\tau=\hbar/\bar E\simeq 9.82~\mu\mathrm{s}$ \\
        \hline
            potential depth & $V_p = 10.0 \bar{E} = 20 E_r$ \\
        \hline
            interaction & $g_0= g_{1D}/(\bar E\bar L)$ \\
        \hline \hline
    \end{tabular}
    \caption{Units used in MCTDH-X simulations. $E_r = \frac{\hbar^2 k_p^2}{2m}$ is the recoil energy of the primary lattice}.
    \vspace{-5pt}
\end{table}

The quasi periodic lattice is the superposition of two optical lattices; a primary lattice of depth $V_p$ and wavelength $\lambda_p$ and a secondary lattice of depth $V_d$ and wavelength $\lambda_d$ parameterized as:
\begin{equation}
    V(x) = V_p\sin^2(k_px) + V_d \sin^2(k_dx),
\end{equation}
where $k_i$ is the wave vector. We use the system parameters with interaction strengths $g_0\in\{0,\,0.001,\,0.05,\,0.5\}$ for the ground states and $g_0\in\{0.05,\,0.5\}$ for the dynamics. We choose $\lambda_p \simeq 532.2$~nm and $\lambda_d \simeq 444.5$~nm, which are compatible with real experimental realizations in ultracold atomic gases. These give vectors $k_p \simeq 1.1806\times10^{7}$~m$^{-1}$ and $k_d \simeq 1.4134\times10^{7}$~m$^{-1}$, with $k_d/k_p\simeq1.197215$. We impose hard-wall boundaries to restrict the optical lattice to the central wells, located at $x=\pm9\pi/2$, and perform no average over the relative phase of the two lattices.

\subsection{Lengths}
In MCTDH-X simulations, we choose to set the unit of length $\bar{L} \equiv 1/k_p=\lambda_p/2\pi \simeq 84.7$~nm, which makes the minima of the primary lattice appear at integer multiples of $\pi$ in dimensionless units, while the maxima are located at half-integer multiples of $\pi$. $x = 0$ is the center of the lattice which can host an odd number of lattice sites $S$. The lattice spacing is $d=\lambda_p/2\simeq266.1$~nm, the nine-site domain has a length of about $2.39~\mu$m, and the spatial grid has 512 points ($\Delta x\simeq4.68$~nm).

\subsection{Energies}
The unit of energy $\bar{E}$ is defined in terms of the recoil energy of the primary lattice, i.e. $E_r \equiv \frac{\hbar^2 k_p^2}{2m} \simeq 5.37\times10^{-30}$~J ($E_r/h\simeq8.11$~kHz) with $m\simeq 86.909$~u, the mass of $^{87}Rb$ atoms. Thus we define the unit of energy as $\bar{E} \equiv \frac{\hbar^2}{m{\bar L}^2} = 2E_r \simeq 1.074\times10^{-29}$~J. In typical experiments with quasi periodic optical lattice, the depth of the primary lattice is varied in the around few tens of recoil energies and the depth of the secondary lattice is varied around few recoil energy. In our simulations, we probe similar regimes: $V_p = 10 {\bar E} = 20E_r$; 
$V_d \in \left[0.5 {\bar E}, \, 5{\bar E}\right]=[1,\,10]\,E_r$. These values are experimentally accessible in ultracold quantum simulators.  The contact coupling is expressed in units of $\hbar^2/(m\bar L)\simeq9.10\times10^{-37}$~J\,m, so that $g_0=0.05$ and $0.5$ correspond to $g_{1D}\simeq4.55\times10^{-38}$ and $4.55\times10^{-37}$~J\,m; the relation to the three-dimensional scattering length depends on the transverse confinement, which is not specified here.

\subsection{Time}
The unit of time is defined from the unit of length as $\bar{t} \equiv \frac{m \bar{L}^2}{\hbar}=\frac{\hbar}{\bar E}$ $\simeq 9.82~\mu\text{s}$. The simulated window, $t\le1000$, thus corresponds to about $9.8$~ms.

\bibliography{ref}

\end{document}